\documentclass[12pt,reqno]{amsart}

\usepackage{amsxtra}
\usepackage{amsfonts}
\usepackage{amsmath,amssymb,stmaryrd} 
\usepackage[english]{babel} 
\usepackage{graphicx} 
\usepackage{mathrsfs} 
\usepackage{layout}
\usepackage{color}

\newcommand{\ba}{\begin{eqnarray}} 
\newcommand{\ea}{\end{eqnarray}}
\newcommand{\be}{\begin{equation}}
 \newcommand{\ee}{\end{equation}}
\newcommand{\bdm}{\begin{displaymath}}
\newcommand{\edm}{\end{displaymath}} 
\newcommand{\brr}{\begin{array}}
\newcommand{\err}{\end{array}}

\newcommand{\bml}{\begin{gather}} \newcommand{\eml}{\end{gather}}

\newcommand{\R}{{\mathbb R}}

\newcommand{\N}{{\mathbb N}}

\newcommand{\var}{\varepsilon}

\newcommand{\pa}{\partial}

\begin{document}

\newtheorem{theorem}{Theorem}[section]
\newtheorem{lemma}[theorem]{Lemma}
\newtheorem{observation}[theorem]{Observation}
\newtheorem{definition}[theorem]{Definition}
\newtheorem{example}[theorem]{Example}
\newtheorem{corollary}[theorem]{Corollary}
\newtheorem{assumption}{Assumption}


\title{ Semiclassical Propagation of Coherent States for the Hartree equation }

\maketitle
\begin{center}
A. Athanassoulis\footnote{CMLS, \'Ecole Polytechnique, France - \emph{e-mail:} agis.athanassoulis@math.polytechnique.fr}, 
T. Paul\footnote{CNRS and CMLS, \'Ecole Polytechnique, France - \emph{e-mail:} thierry.paul@math.polytechnique.fr},
F. Pezzotti\footnote{Departamento de Matem\'aticas, Universidad del Pa\'is Vasco, Spain - \emph{e-mail:} federica.pezzotti@ehu.es},
M. Pulvirenti\footnote{Dipartimento di Matematica ``G. Castelnuovo'', Universit$\grave{\text{a}}$ di Roma ``La Sapienza'', Italy - \emph{e-mail:} pulvirenti@mat.uniroma1.it }
\end{center}


\tableofcontents




\section{Introduction}

Let us consider the Hartree equation in $\Bbb R^d$:
\begin{equation}
\label{eq1}
\begin{array}{c}
i \var \pa_t \Psi ^\var (x,t)= -\frac {\var^2}2 \Delta \Psi ^\var (x,t)+\left(V(x,t)+U(x,t)\right) \Psi ^\var (x,t), \\ { } \\
\Psi^\var(x,0)=\Psi^\var_0(x),
\end{array}
\end{equation}
where
\begin{equation}
\label{eqV}
V(x,t)=\int \phi (\vert x-y\vert) |\Psi ^\var (y,t)|^2dy
\end{equation}
is a self-consistent potential given by a smooth two-body interaction, $\phi:\mathbb{R} \shortrightarrow \mathbb{R}$, even, and $U(\cdot,t):\mathbb{R}^d \shortrightarrow \mathbb{R}$ for all $t\geq 0$, is a smooth external potential 
(see the next section for the precise assumptions on $\phi$ and $U$). \\

 In a recent paper \cite{appp1} the authors of the present one considered the semiclassical limit 
 of the version of the Hartree equation corresponding to mixed states, for initial data whose Wigner functions do not
 concentrate at the classical limit.
 
The problem we deal with  in the present paper is the   semiclassical asymptotics  for (\ref{eq1}) when the initial state is a coherent state centered around the point $q,p$ of the classical phase space, namely:
\begin{equation}
\label{eqdefchst}
 \Psi ^\var_0 (x)=\var ^{-\frac d 4} a_0\left(\frac {x-q}{\sqrt {\var}}\right) e^{i \frac {p \cdot (x-q)}{\var}}:=\psi^{a_0}_{qp}(x).
\end{equation}

\vskip 0.25cm
This problem was studied  in 
\cite{LP} in the kinetic (Wigner) picture, 
see  Th\'eor\`eme IV.2 therein. There it is shown that, under appropriate conditions, the solution $W^\var$ of the Wigner equation corresponding to the dynamics (\ref{eq1})
%
namely
\begin{eqnarray}
 \partial_t W^\var + k \cdot \partial_x W^\var &=&\frac{i}{\var(2\pi)^d} \int \int e^{i\xi y}\left(V(x+\frac{\var y}{2},t)-V(x-\frac{\var y}{2},t)
\right)dy  \, W^\var(x,k-\xi)d\xi +\nonumber\\
& + &\frac{i}{\var(2\pi)^d} \int \int e^{i\xi y}\left(U(x+\frac{\var y}{2},t)-U(x-\frac{\var y}{2},t)
\right)dy  \, W^\var(x,k-\xi)d\xi,
\end{eqnarray}
where $V(x,t)$ is the same as in (\ref{eq1}) equivalently written as
\begin{equation}
\label{eqVV}
V(x,t) = \int{ \phi(\vert x-y\vert ) W^\var(y,k,t)dk dy},
\end{equation}
converges, in weak$*$-sense, to the solution of the (classical) Vlasov equation
\begin{equation}
\label{Vlasov}
\begin{array}{c}
\partial_t f + k \cdot \partial_x f- \partial_x V_0(x,t) \cdot \partial_k f-\partial_x U(x,t) \cdot \partial_k f=0, \\ { } \\
f(x,k,t)\vert_{t=0}=f_0(x,k),
\end{array}
\end{equation}
where 
$$
V_0(x,t)=\int{ \phi(\vert x-y\vert) f(y,k,t)dk dy},
$$
and $U(x,t)$ is  the same as in (\ref{eq1}) . The initial condition for (\ref{Vlasov}) is given by $f_0=w-^{*} \lim\limits_{\var \to 0} W^\var_0$. It is easy to check that the conditions of  
Th\'eor\`eme IV.2 in \cite{LP} are satisfied for $W^\varepsilon_0(x,v)=W^\varepsilon[\Psi^\var_0](x,v)$, $\Psi^\var_0$ as in equation (\ref{eqdefchst}). In that case (under appropriate assumptions 
on the pair-interaction potential $\phi$ and the external potential $U$) it can be seen that the Wigner measure of the wave function verifies
\[
W^\var[\Psi^\var](x,k, t) \,\, \rightharpoonup \,\, \delta(x-X(t))\delta(k-K(t)),\ \ \ \ \text{as}\ \, \var\to 0,
\]
where
\[
\begin{array}{c}
\dot X(t) =K(t), \,\,\,\,\,\,\,\,\,\, \dot K(t)=-\nabla U(X(t),t),  \ \ \ \ \ \ \ \ \ 
X(0)=q, \,\,\,\,\,\,\,\,\,\, K(0)=p.
\end{array}
\]
In that sense,  the semiclassical limit of the problem (\ref{eq1}) is known to be the Vlasov dynamics (\ref{Vlasov}), 
since it is easy to recognize that, due to the smoothness of the potentials, the limiting measure 
$\delta(x-X(t))\delta(k-K(t))$  is the unique (weak) solution of the Vlasov equation with initial datum $\delta(x-q)\delta(k-p)$.

The goal of the present work is to strengthen this approximation. First of all, we construct $L^2$ 
approximations, as opposed to the with weak$*$-limit, and this yields an explicit control of the error in $\var$ which allows to recover the shape with which $W^\var$ 
concentrates to a $\delta$ in phase-space. 

\section{Main result}
We will consider the Hartree equation in $\Bbb R^d$:
\begin{equation}
\label{1eq1}
\begin{array}{c}
i \var \pa_t \Psi ^\var (x,t)= -\frac {\var^2}2 \Delta \Psi ^\var (x,t)+\left(V(x,t)+U(x,t)\right) \Psi ^\var (x,t), \\ { } \\
\Psi^\var(x,0)=\Psi^\var_0(x),
\end{array}
\end{equation}
where
\begin{equation}
\label{1eqV}
V(x,t)=\int \phi (\vert x-y\vert) |\Psi ^\var (y,t)|^2dy
\end{equation}

The initial condition will be of the form
\[
 \Psi ^\var_0 (x)=\var ^{-\frac d 4} a_0\left(\frac {x-q}{\sqrt {\var}}\right) e^{i \frac {p \cdot (x-q)}{\var}}:=\psi^{a_0}_{qp}
\]
and we will make the following assumptions on $a_0,\ \phi$ and $U$:\vskip 0.3cm
\begin{assumption} \label{A1}
 $$
\left\|a_0\right\|_{L^2}=\left\|\Psi_0^\var\right\|_{L^2}=1,
$$

$$x^A  \partial_x^B  a_0(x)  \in L^2\mbox{ for any  pair } A,B  \in \mathbb{N}^d\mbox{ with }|A|+|B| \leqslant 3,$$
\begin{eqnarray}
&&\int x_i \vert a_0(x)\vert^2 d x =0,\  \forall\,  i=1\dots d.\label{theta}\\
&&\int k_i \vert \widehat{a}_0(k)\vert^2 d k =0,\  \forall\,  i=1\dots d.\label{thetak}
\end{eqnarray}
\end{assumption}
\begin{assumption} \label{A2}

$$ 
C^3_b(\R)\ni\phi \ even
$$ 

\end{assumption}

\begin{assumption} \label{A3}

$$U \in C^1\left(\R^+_t, C^3_b(\R^d_x)\right).$$ 
\end{assumption}
\vskip 0.3cm
Here and henceforth we denote by $C_b^k(\R^m)$ the space of continuous and uniformly 
bounded functions on $\R^m$ whose all derivatives up the order $k$ are also  continuous and uniformly bounded.

\begin{theorem} \label{thrm1} 
Under Assumptions \ref{A1}, \ref{A2} and \ref{A3} 
there exists a 
constant $C$ such that, $\forall\  t\geq 0$, 
\begin{equation}\label{MAINclaim}
\Vert \Psi^\var(\cdot,t)-e^{i\frac{\mathcal{L}(t)}{\var}+i\gamma(t)}\psi^{\beta_t}_{q(t)p(t)}\Vert_{L^2} 
\leqslant 
Ce^{ C t  e^{C\,  e^{C\,  t}}  }\cdot \sqrt{\varepsilon}.
\end{equation}
where $\beta_t$ is the solution of
\begin{equation}
\label{machin}
i\partial_t\beta_t(x)=\left(-\frac{\Delta}{2}+\frac{\phi''(0) x^2}{2}+\frac{<x,\nabla^2 U(q(t),t)x>}2\right)\beta_t(x),
\end{equation}
\begin{equation}\label{cxz}
\beta_0(x)=a_0(x),\end{equation}

\begin{equation}\label{vcx}
\gamma(t)=-\frac{\phi''(0)}{2}\int_0^t\int\eta^2\vert\beta_s(\eta)\vert^2d\eta ds,
\end{equation}

$(q(t),p(t))$ is the Hamiltonian flow associated with $\frac{p^2}2+U(q,t)+\phi(0)$ issued from $(q,p)$, 

$$\mathcal{L}(t):=\int_0^t\left(p(s)^2/2-U(q(s),s)-\phi(0)\right)ds$$ 

\centerline{(the Lagrangian action along such Hamiltonian flow).}



\end{theorem}

\vspace{0.5cm}

\noindent {\bf Remarks:} 

 
\begin{itemize}
\item
As shown in the proof of the Theorem, the constant $C$ depends 
only on $d$, $||U||_{W^{3,\infty}}$, $||\phi||_{W^{3,\infty}}$ and $\mathop{sup}\limits_{|A|+|B|\leqslant 3} ||x^B \partial_x^A a_0||_{L^2}$. 

\item  Note that in the classical flow the nonlinear potential enters only via the inessential constant $\phi (0)$. Indeed, due to the 
symmetry and smoothness of $\phi$, we have $ \phi '(0)=0$ so that, in case of concentration as $\var\to 0$, the self-consistent field $\nabla V$
vanishes.

\item A similar  problem  for $\phi'' (0) \geq 0$ has been faced in \cite {Bel2} in a semirigorous way. Here we treat the case 
$\phi '' (0) \leq 0$ as well and present an explicit control of momenta and derivatives of the solutions (see Lemma 2.3 below)
which allow us to estimate the error in $L^2$.
\item For a related result (Gross-Pitaevskii equation with a different scaling)
see \cite{carlito}.


\item
Assumption \ref{A1} can be relaxed by dismissing equation (\ref{theta}). Indeed even if (\ref{theta}) does not hold one can always make a change of variables $x \mapsto x-\int x \vert a_0(x)\vert^2 d x$. However in that case one would have to adjust appropriately the external potential, which of course is not translation invariant.
\end{itemize} 
\vskip 0.2cm



\section{Proofs}

\subsection{A Lemma}

 We first prove the following
\begin{lemma}\label{14juillet}
$b_t(x):=e^{i\gamma(t)}\beta_t(x)$ as defined by (\ref{machin},\ref{cxz},\ref{vcx}) is the unique solution of  the equation:
\begin{equation}\label{truc}
 \begin{array}{c}
\left({ i  \pa_t +\frac {1}2 \Delta }\right) b_t(x)= \frac{\phi''(0)}{2} \int{ |x-\eta|^2 |b_t(\eta)|^2 d\eta} \,\,b_t (x)
+\frac{<x,\nabla^2 U(q(t),t)x>}2 b_t (x), \\ { } \\
 b_0(x)=a_0(x).
\end{array}
\end{equation}

\end{lemma}
\begin{proof}\rm
\begin{equation}\label{derB}
 i\pa_t b_t(x)=-\gamma'(t) b_t(x) +  e^{i\gamma(t)}i\pa_t\beta_t(x).
\end{equation}
By virtue of equations (\ref{machin}), (\ref{cxz}) and (\ref{vcx}) we find
\begin{equation}\label{Beqn0} 
 \begin{array}{c}
i\pa_t b_t(x)=\frac{\phi''(0)}{2}\int\eta^2\vert\beta_t(\eta)\vert^2d\eta\ b_t(x) +e^{i\gamma(t)}\left(-\frac{\Delta}{2}\beta_t(x)+\frac{\phi''(0)}{2}x^2\ \beta_t(x)+\frac{<x,\nabla^2 U(q(t),t)x>}2\ \beta_t(x) \right)
, \\ { } \\
 b_0(x)=a_0(x),
\end{array}
\end{equation}
namely
\begin{equation}\label{Beqn}
 \begin{array}{c}
i\pa_t b_t(x)=-\frac{\Delta}{2}b_t(x)+\frac{\phi''(0)}{2}x^2\ b_t(x)+\frac{\phi''(0)}{2}\int\eta^2\vert\beta_t(\eta)\vert^2d\eta\ b_t(x) +\frac{<x,\nabla^2 U(q(t),t)x>}2\ b_t(x) 
, \\ { } \\
 b_0(x)=a_0(x).
\end{array}
\end{equation}
 
We first notice that the equation (\ref{machin}) for $\beta_t(x)$ is  a linear Schr\"odinger equation with an harmonic potential; therefore the solution $\beta_t(x)$ of the initial value problem (\ref{machin})-(\ref{cxz}) is uniquely determined in $L^2(\R^d)$ and
\begin{equation}\label{l2beta}
\left\|\beta_t\right\|_{L^2}=\left\|a_0\right\|_{L^2}=1,\ \ \forall\ t\in\R.
\end{equation}
As a consequence of that, it turns out that equation (\ref{Beqn}) can be rewritten as
\begin{equation}\label{Beqn1}
 \begin{array}{c}
i\pa_t b_t(x)=-\frac{\Delta}{2}b_t(x)+\frac{\phi''(0)}{2}\int x^2\vert\beta_t(\eta)\vert^2 d\eta \ b_t(x)+\frac{\phi''(0)}{2}\int\eta^2\vert\beta_t(\eta)\vert^2d\eta\ b_t(x) +\frac{<x,\nabla^2 U(q(t),t)x>}2\ b_t(x) 
, \\ { } \\
 b_0(x)=a_0(x).
\end{array}
\end{equation}
Furthermore, it is easy to check that if
\begin{equation}\label{MOMa}
x a_0(x), \,\, \partial_x a_0(x) \in L^2(\R^d),
\end{equation}
then
\begin{equation}\label{MOMbeta}
x \beta_t(x), \,\, \partial_x \beta_t(x) \in L^2(\R^d),\ 
\text{for all}\ t,
\end{equation}
(see Observation \ref{obsobsobsBETA} below).\\
Condition (\ref{MOMa}) is satisfied under Assumption \ref{A1}, so the property (\ref{MOMbeta}) holds 
and, in particular, there exists a constant $C$ finite for any time $t$ 
such that
\begin{equation}\label{MOMbetaPROD}
\int \vert \eta \vert^2 \vert \beta_t(\eta)\vert^2 d \eta<C, \,\, \forall\ \ t\in\R.
\end{equation}
Thus, by virtue of (\ref{MOMbetaPROD}) and of Assumptions \ref{A1}, \ref{A2} and \ref{A3}, it follows that the initial value problem (\ref{Beqn1}) is guaranteed to have a unique solution in $L^2$ and, clearly, $\left\|b_t\right\|_{L^2}=\left\|a_0\right\|_{L^2}=1,\ \ \forall\ t$. In fact, the equation for $b_t(x)$ has turned to be a linear Schr\"odinger equation with an harmonic potential (and all constants appearing in the potential terms are finite thanks to Assumptions \ref{A2} and \ref{A3} and to (\ref{MOMbetaPROD})).

Now, it  remains only to recognize that (\ref{Beqn1}) is exactly the same as (\ref{truc}). To this end it is sufficient to observe that, since the equation (\ref{machin}) for $\beta_t(x)$ is a linear Schr\"odinger equation with an harmonic potential and conditions (\ref{theta}) and (\ref{thetak}) are satisfied at time $t=0$, we are guaranteed that
\begin{equation}\label{MOM1beta}
\int \eta \vert \beta_t(\eta)\vert^2 d \eta=0, \,\, \forall\ \ t.
\end{equation}
Thus, by virtue of (\ref{MOM1beta}), it follows straightforwardly that (\ref{Beqn1}) can be rewritten as
\begin{equation}\label{Beqn2}
 \begin{array}{c}
i\pa_t b_t(x)=-\frac{\Delta}{2}b_t(x)+\frac{\phi''(0)}{2}\int \vert x-\eta\vert^2\vert\beta_t(\eta)\vert^2 d\eta \ b_t(x) +\frac{<x,\nabla^2 U(q(t),t)x>}2\ b_t(x) 
, \\ { } \\
 b_0(x)=a_0(x).
\end{array}
\end{equation}
Finally, it is clear, by the definition of $b_t(x)$, that $\vert \beta_t(x)\vert=\vert b_t(x)\vert$ for any $x$ and $t$. Therefore (\ref{Beqn2}) turns to be exactly the same as (\ref{truc}).
\end{proof}


\subsection{Proof of Theorem \ref{thrm1}}
We seek an approximate solution to equation (\ref{eq1}) of the form as e.g. in \cite{HAG1,HAG,HEPP,tp1,tp2} 
\begin{equation}
\label{eqans}
 \Psi ^\var (x,t)=\var ^{-\frac d 4} a (\frac {x-q(t)}{\sqrt {\var}},t ) e^{i \frac {p(t) \cdot (x-q(t))}{\var}}
 e^{i \frac {\mathcal{L}(t) }{\var}}  
\end{equation} 
where
\begin{equation}
\dot q(t)=p(t),\ \dot p(t)=-\nabla U(q(t),t).
\end{equation}

By inserting the ansatz (\ref{eqans}) in equation (\ref{eq1}) we get
\begin{eqnarray}\label{ansatz1}
i \var \pa_t \Psi ^\var (x,t)&=&\var ^{-\frac d 4} \left[i \var \pa_t a(\frac {x-q(t)}{\sqrt {\var}},t )-i  \sqrt {\var} \nabla a(\frac {x-q(t)}{\sqrt {\var}},t ) \cdot  {\dot q(t)}+ \right.\nonumber\\
&&- \left.  \mathcal{L}'(t)a(\frac {x-q(t)}{\sqrt {\var}},t )-\left(\dot{p}(t)(x-q(t))-p(t)\dot{q}(t)\right)a(\frac {x-q(t)}{\sqrt {\var}},t )\right]\times\nonumber\\
&&\times
e^{ i\frac {p(t) \cdot (x-q(t))}{\var}}
e^{ i\frac {\mathcal{L}(t) }{\var}} ,
\end{eqnarray}
and
\begin{eqnarray}\label{ansatz2}
-\frac {\var^2}2 \Delta \Psi ^\var (x,t)&=&\var ^{-\frac d 4}\left[ -\frac {\var}2 \Delta a(\frac {x-q(t)}{\sqrt {\var}},t)+\frac {p^2(t)}2 a(\frac {x-q(t)}{\sqrt {\var}},t)+\right.\nonumber\\
&&\left.-i  \sqrt {\var} \nabla a(\frac {x-q(t)}{\sqrt {\var}},t)  \cdot  {p}(t)\right]
e^{ i\frac {p(t) \cdot (x-q(t))}{\var}}
e^{ i\frac {\mathcal{L}(t) }{\var}},
\end{eqnarray}
while, with regard to the potential terms in (\ref{eq1}), we find
\begin{eqnarray}\label{ansatz3}
 \left(V(x,t)+U(x,t)\right)\Psi ^\var (x,t)&=&\var^{-d/4}\left(\int \phi (\vert x-y\vert )\var^{-d/2} |a (\frac {y-q(t)}{\sqrt {\var}},t)|^2 dy+U(x,t)\right)\times\nonumber\\
 && \times\,  a (\frac {x-q(t)}{\sqrt {\var}},t)e^{ i\frac {p(t) \cdot (x-q(t))}{\var}}
e^{ i\frac {\mathcal{L}(t) }{\var}}.
\end{eqnarray}
By (\ref{ansatz1}), (\ref{ansatz2}) and (\ref{ansatz3}) we get that the amplitude $a$ solves the following initial value problem:
\begin{equation}
\label{eq1sw}
\begin{array}{c}
\left({ i \pa_t +\frac {1}2 \Delta }\right) a (\mu,t)=\frac 1 \var V_\var (\mu,t)a 
(\mu,t)+\\ 
{ } \\+\frac 1 \var\left[U(q(t)
+\sqrt\var\mu,t)-U(q(t),t)-\sqrt\var \nabla U(q(t),t)\cdot\mu\right]a (\mu,t), 
\\ { } \\
a(\mu,0)=a_0(\mu),
\end{array}
\end{equation}
where
\begin{equation}
\label{eqeag}
V_\var (\mu,t)= \int   \left({ \phi  (\sqrt{\var} \vert \mu-\eta\vert) -\phi (0) }\right)   |a(\eta,t)|^2 d \eta,
\end{equation}
$q(t),p(t)$ are as in the claim of Theorem \ref{thrm1} and we have used the rescaling $\mu=\frac{x-q(t)}{\sqrt{\varepsilon}}$.\\
Note that we should have
\begin{equation}
\label{eqeag1}
V_\var (\mu,t)= \int    \phi\left({ \sqrt{\var} \vert \mu-\eta\vert}\right) |a(\eta,t)|^2  d \eta  -\phi (0) ,
\end{equation}
instead of (\ref {eqeag}) in equation (\ref{eq1sw}). However equation  (\ref{eq1sw}) with potential (\ref {eqeag1}) is an Hartree equation
which preserves the $L^2$ norm  so that we can replace (\ref {eqeag1}) by (\ref {eqeag}).


Since $\phi \in C_b^3(\R)$ is even $\phi'  (0)=0$ and  the Taylor expansion yields
\begin{equation}\label{phiAPPROX}
\begin{array}{c}
\phi(\sqrt{\var}\vert \mu-\eta \vert)-\phi(0)=\frac{\var |\mu-\eta|^2}{2}\phi''(0)+ \var^{\frac{3}2} R(|\mu-\eta|), \\ { } \\

|R(|\mu-\eta|)| \leqslant C ||\phi'''||_{L^\infty} |\mu-\eta|^3,
\end{array}
\end{equation}
while for the terms involving $U$ we find:
\begin{equation}\label{uAPPROX}
\begin{array}{c}
U(q(t)+\sqrt\var\mu,t)-U(q(t),t)-\sqrt\var \nabla U(q(t),t)\cdot \mu = \var<\mu,\frac{\nabla^2 U(q(t),t)}2\mu>+ \var^{\frac{3}2} R_U(\mu,t), \\ { } \\
|R_U(\mu,t)| \leqslant C\,  \sup_{\alpha\in\N^d:\vert \alpha\vert=3}|\nabla^\alpha U(q(t),t)| |\mu|^3,
\end{array}
\end{equation}
where $\nabla^2:=\nabla\otimes\nabla$.

The core of the proof is to estimates the two remainders $\var^{\frac{3}2}R(|\mu-\eta|) $ and 
$\var^{\frac{3}2} R_U(\mu,t)$ so that we can substitute $\left(\phi(\sqrt{\var}\vert \mu-\eta\vert)-\phi(0)\right)$ by $\frac{\var |\mu-\eta|^2}{2}\phi''(0)$ (as in (\ref{phiAPPROX})) and $U(q(t)+\sqrt\var\mu,t)-U(q(t),t)-\sqrt\var \nabla U(q(t),t)\cdot \mu$ by $\var<\mu,\frac{\nabla^2U(q(t),t)}2\mu>$ (as in (\ref{uAPPROX})). \\
 
In the framework of semiclassical approximation for the linear Schr\"odinger equation using coherent states
 the method is standard (see e.g.  \cite{HAG1, HAG,HEPP, tp1,tp2}), 
 however we establish these estimates again
 for completeness.

Denote $a_t(\mu):=a(\mu,t)$ and
\begin{equation}
h_t(\mu)=b_t(\mu)-a_t(\mu).
\end{equation}
By straightforward substitution we get that $h_0(\mu)=0$ (see (\ref{truc})) and
\begin{eqnarray}
\label{eqr}
&& \left(i  \pa_t +\frac {1}2 \Delta-\underbrace{\left(\frac{\phi''(0)}{2} \int{ |\mu-\eta|^2 |b_t(\eta)|^2 d\eta}+<\mu,\frac{\nabla^2 U(q(t),t)}2\mu>\right)}_{V_Q(\mu,t)}\right) h_t (\mu)=  \nonumber\\ 
&& = \frac{\phi''(0)}{2} \underbrace{\int{ |\mu-\eta|^2 \left({ |b_t(\eta)|^2-|a_t(\eta)|^2 }\right) d\eta} \,\,a_t(\mu)}_{I_1(\mu,t)}  + \nonumber\\
&& - \sqrt{\varepsilon}  \underbrace{\int{ R(|\mu-\eta|) |a_t(\eta)|^2 d\eta} \,\, a_t(\mu)}_{I_2(\mu,t)} -\sqrt{\var} R_U(\mu,t)a_t(\mu).
\end{eqnarray}

By standard manipulations 
it turns out that
\begin{equation}
\label{eqr1}
\left\|h_t\right\|_{L^2} \frac{d}{dt}\left\|h_t\right\|_{L^2} \leqslant \frac{|\phi''(0)|}{2} {  |\langle I_1,h_t \rangle| + \sqrt{\var} |\langle I_2,h_t \rangle| +\sqrt{\var}|\langle R_U(\cdot, t) a_t,h_t \rangle|}.
\end{equation}
Moreover, the term involving $I_1$ can be estimated as follows:
\begin{eqnarray}\label{35}
|\langle I_1,h \rangle| &\leq& \left|{ \int\limits_{\mu}{ \int\limits_{\eta} {|\mu-\eta|^2 \left({ |b_t(\eta)|^2-|a_t(\eta)|^2 }\right) d\eta} \,\, \overline{a}_t(\mu)h_t(\mu)   d\mu} }\right| =\nonumber\\
&=&\left|{ \int\limits_{\mu}{ \int\limits_{\eta} {|\mu-\eta|^2 \left({ |b_t(\eta)| -|a_t(\eta)|}\right) \left({|b_t(\eta)| +|a_t(\eta)| }\right)
 d\eta} \,\, \overline{a}_t(\mu)h_t(\mu)   d\mu} }\right| \leq \nonumber\\
&\leq&  \left|{ \int\limits_{\mu}{ \int\limits_{\eta} {|\mu-\eta|^2 |h_t(\eta)| \left({|b_t(\eta)| +|a_t(\eta)| }\right) d\eta} \,\, \overline{a}_t(\mu)h_t(\mu)   d\mu} }\right| \leq \nonumber\\
&\leq& 2\left\|h_t\right\|_{L^2}^2 \int(1+|\mu|^2)^2\left[ |a_t(\mu)|+|b_t(\mu)| \right]^2 d\mu,
\end{eqnarray}
while, thanks to (\ref{phiAPPROX}), the term involving $I_2$ is estimated by:
\begin{equation}\label{36}
\begin{array}{c}
|\langle I_2,h \rangle| \leq C\left\|\phi'''\right\|_{L^\infty} \left|{ \int\limits_{\mu}{ \int\limits_{\eta} {|\mu-\eta|^3 |a_t(\eta)|^2 d\eta} \,\, \overline{a}_t(\mu)h(\mu,t)   d\mu} }\right| \leq \\ { } \\

\leq C \left\|\phi'''\right\|_{L^\infty}\left({  
\left(\int d\eta\, \vert \eta\vert^3\vert a_t(\eta)\vert^2\right) ||a_t||_{L^2}||h_t||_{L^2}+ }\right. \\ { } \\

\left.{ + 3\left(\int d\eta\, \vert \eta\vert^2\vert a_t(\eta)\vert^2\right)^{3/2}
||h_t||_{L^2}+
3\left(\int d\mu\, \vert \mu\vert^4\vert a_t(\mu)\vert^2 \right)^{1/2}\, \left(\int d\eta\, \vert \eta\vert\, \vert a_t(\eta)\vert^2\right)
||h_t||_{L^2} + }\right. \\ { } \\

\left.{+\left(\int d\mu\, \vert \mu\vert^6\vert a_t(\mu)\vert^2 \right)^{1/2}|| a_t||_{L^2}^2
||h_t||_{L^2}
 }\right).
\end{array}
\end{equation}
One should observe here that $\int d\eta\, \vert \eta\vert\, \vert a_t(\eta)\vert^2\leq \left(\int d\eta\, (1+\vert \eta\vert^2)\, \vert a_t(\eta)\vert^2\right)$.\\

Finally, due to (\ref{uAPPROX}), the term involving $R_U(\mu,t)$ is controlled as follows:
\begin{eqnarray}\label{RUcontrol}
|\langle R_U(\cdot, t) a_t,h_t \rangle|&\leq &C\, \sup_{\alpha:\vert \alpha\vert=3} \left| \nabla^\alpha U(q(t),t)\right| \left(\int d\mu\, \vert \mu\vert^3\vert a_t(\mu)\vert\, \vert h_t(\mu)\vert \right)\leq \nonumber\\
&\leq & C\,  \sup_t \sup_{\alpha:\vert \alpha\vert=3}\left| \nabla^\alpha U(q(t),t)\right|\left(\int d\mu\, \vert \mu\vert^6\vert a_t(\mu)\vert^2 \right)^{1/2}||h_t||_{L^2}.
\end{eqnarray}

\vskip 0.4cm
Making use of Lemma \ref{lmmom} and equation (\ref{b3CONTROL}) below to estimate terms 
of the form $||\,\,|\cdot|^m a_t||_{L^2}=\left(\int d\eta\vert \eta\vert^{2m}\vert a_t(\eta)\vert^2\right)^{1/2}$, 
for $m\leq 3$, and $||\,\,|\cdot|^m b_t||_{L^2}=\left(\int d\eta\vert \eta\vert^{2m}\vert b_t(\eta)\vert^2\right)^{1/2}$, 
for $m\leq 2$, in terms of the same quantities evaluated at time $t=0$, we easily show, 
by summing up the previous estimates, that there exist three $\var$-independent functions $C_1(t), C_2(t)$
such that: 
\begin{equation}
\frac{d}{dt}||h_t||_{L^2} \leq  \sqrt{\varepsilon} C_1(t) + C_2(t) ||h_t||_{L^2}.
\end{equation}
In particular $C_1(t), C_2(t)$ depend on the potentials $\phi$ and $U$ and on the $L^2$-norm of moments and derivatives of $a_0$ (up to the order $3$). With regard to the time dependence, $C_1(t),C_2(t)$ are double exponentials $Ce^{Ce^{Ct}}$, following Lemma \ref{lmmom} and observations \ref{obsobsobsBETA}, \ref{obsobsobs}.

The conclusion follows with application of the Gronwall lemma.

\hfill $\square$ 

\section{Auxiliary results}

\begin{observation} \label{obsprop} Observe that under our assumptions the nonlinear equation  (\ref{eq1sw}) can be shown to have, for any $T>0$, a unique solution in
 $C^1\left({ [0,T], L^2(\mathbb{R}^d) }\right)$ (see e.g. \cite{CW}). Therefore it follows  (see e.g. \cite{SR}) that the corresponding {\em time-dependent linear} problem
\begin{equation}
\label{eq1swclo}
\begin{array}{c}
\left({ i \pa_t +\frac {1}2 \Delta }\right) u (\mu,t)=\frac{1}{\var}\int   \left({ \phi ( \sqrt{\var} |\mu-\eta|) -\phi (0) }\right)   |a(\eta,t)|^2 d \eta \,\, u(\mu,t)+\\ { } \\
\ \ \ \ +\frac 1 \var\left(U(q(t)+\sqrt\var\mu,t)-U(q(t),t)-\sqrt\var \nabla U(q(t),t)\cdot\mu\right)u (\mu,t), \\ { } \\
u(x,0)=u_0(x),\ \ \ u_0\in L^2(\R^d)\, \ \left\|u_0\right\|_{L^2}=1,
\end{array}
\end{equation}
has a unique and well-defined $L^2$ propagator.
\end{observation}
\vskip 0.3cm

\begin{lemma}[Propagation of Moments and derivatives for $a(x,t)$] \label{lmmom} 

Let $a(x,t)$ be the solution of the initial value problem (\ref{eq1sw}). 
Suppose that for some $m\in \mathbb{N}$ there exists an $\var$-independent constant $M_m>0$ such that
\begin{equation}\label{hypLEMMAa}
 || x^A \partial_x^B a_0||_{L^2} \leqslant M_m
\end{equation}
for all $A, B \in \mathbb{N}^d$ such that $|A|+|B|\leqslant m$.

For $m\geq 2$, assume $\phi\in C_b^{m}(\R^d)$  
and $U\in C^{1 }(\R^+_t, C_b^{m}(\R_x^d))$. Then, there exists a (finite) $\var$-independent constant $C_m$  such that 
\begin{equation}\label{momA}
 || x^A \partial_x^B a(t)||_{L^2} \leqslant C_m e^{C_m e^{C_m t}} M_m,
\end{equation}
for all $A, B \in \mathbb{N}^d$ such that $|A|+|B|\leqslant m$.

For $m=1$ inequality (\ref{momA}) holds by assuming $\phi\in C_b^{2}(\R^d)$  
and $U\in C^{1 }(\R^+_t, C_b^{2}(\R_x^d))$, while in the case $m=0$ formula (\ref{momA}) becomes an equality and holds with unitary constant (for all $t$) by simply assuming $\phi\in C_b^{0}(\R^d)$ and $U\in C^{1 }(\R^+_t, C_b^{1}(\R_x^d))$.

\end{lemma}

\noindent {\bf Remark:}The proof makes no use of an energy conservation argument,  and this is the reason why the Lemma can be established for  both signs of $\phi''(0)$. 
 
%
%
\vskip 0.2cm

\begin{proof}  

Denote
\begin{equation}
\psi^{A,B}(x,t)=x^B\partial_x^A a(x,t),
\end{equation}
e.g. $\psi^{0,0}(x,t):=a(x,t)$.\\
It is straightforward to check that
\begin{equation}
\label{eqpab}
\begin{array}{r}
{\left({ i\partial_{t}  + \frac{1}{2}\Delta  -\frac{1}{\var}V_\var(x,t)-\frac{1}{\var} U(q(t)+\sqrt{\var}x,t) +\frac{1}{\var}U(q(t),t)+\frac{1}{\sqrt\var} \nabla U(q(t),t)\cdot x}\right) \psi^{A,B}(x,t)} =\ \ \ \ \ \ \ \ \ \ \ \ \ \ \ \ \ \ \ \ \ \ \ \ \ \ \ \ \ \ \ \  \ \ \ \ \ \ \ \ \ \ \ \ \ \ \ \\ { } \\

= {-\sum\limits_{k=1}^{d}  
\left[{ \frac{B_k(B_k-1)}{2} \psi^{A,B-2e_k}(x,t) + B_k\psi^{A+e_k,B-e_k} (x,t)}\right]} + \frac{1}{\var}\sum\limits_{0 \leqslant l < A} { \prod\limits_{k=1}^d \binom{A_k}{l_k} \partial_x^{A-l}V_\var(x,t)  \psi^{l,B}(x,t) +}\ \ \ \ \ \ \ \ \ \ \ \ \ \ \ \  \ \ \ \ \ \ \ \ \ \ \ \ \ \ \ \ \ \  \ \ \ \ \ \ \ \\ { } \\

{+\frac{1}{\var}\sum\limits_{0 \leqslant l < A} { \prod\limits_{k=1}^d \binom{A_k}{l_k} \partial_x^{A-l}U(q(t)+\sqrt{\var}x,t)  \psi^{l,B}(x,t)}-\frac{1}{\sqrt{\var}}\sum\limits_{\substack{0 < l \leqslant A,\\\vert l\vert=1}}
{ \prod\limits_{k=1}^d \binom{A_k}{l_k}\partial_x^{l}\left( \nabla U(q(t),t)\cdot x\right)\psi^{A-l,B}(x,t)}}\ \ \ \ \ \ \ \ \ \ \ \ \ \ \ \ \ \ \ \  \ \ \ \ \ \ \ \ \ \ \ \ \ \ \ \  \ \ \ \ 
\end{array}
\end{equation}
where $B=(B_1,\dots,B_k,\dots,B_d)$, $l=(l_1, \dots,l_k,\dots,l_d)$, $A=(A_1,\dots,A_k,\dots, A_d)$ and $0\leq l<A$ means that $0\leq l_k < A_k$ for any $k=1,2,\dots, d$. The consistent initial data for (\ref{eqpab}) are defined by 
$$
\psi^{A,B}(x,0)=x^B\partial_x^A a_0(x),
$$
and in particular $\psi^{0,0}(x,0):=a_0(x)$.

Some remarks with regard to our notation are in order; it is clear for example that if $B_k=0$ or $B_k=1$, then the first term on the right-hand side yields no contribution. Similarly for $B_k=0$ in the second term and $|A|=0$ for the remaining terms respectively. 	\\
The derivation of (\ref{eqpab}) is straightforward by induction.

Denote by $P(t,\tau)$ the propagator associated with the left-hand side of equation (\ref{eqpab}), which is known to be uniquely well defined in $L^2$ (see Observation \ref{obsprop}). 
As a consequence, for $m=0$, the result claimed by Lemma \ref{lmmom} follows from the existence of the propagator.
We will proceed for $m \in \mathbb{N}$ by induction.

\vskip 0.2cm
We will work with vectors including all the moments and derivatives,  namely, $\overrightarrow{\Psi}=\{\psi^{A,B}\}_{A,B: \vert A\vert + \vert B\vert\leq m}\in X_m$ and
\[
||\overrightarrow{\Psi}||_{X_m}=\sum\limits_{0\leqslant |A|+|B| \leqslant m} ||\psi^{A,B}||_{L^2},
\]
where $X_0:=L^2(\R^d)$.

For $m=1$ we have
\begin{eqnarray}
\label{eq271}
&&\left( i\partial_{t}  + \frac{1}{2}\Delta  -\frac{1}{\var}V_\var(x,t)-\frac{1}{\var} U(q(t)+\sqrt{\var}x,t) +\frac{1}{\var}U(q(t),t)+\frac{1}{\sqrt\var} \nabla U(q(t),t)\cdot x\right) \psi^{e_j,0}(x,t) =\nonumber  \\ 
&&\nonumber\\ 
&&=\frac{1}{\var}\, \partial_{x_j} V_\var(x,t) \psi^{0,0}(x,t)+\frac{1}{\var}\, \partial_{x_j} U(q(t)+\sqrt{\var}x,t) \psi^{0,0}(x,t)-\frac{1}{\sqrt\var}\,  \pa_{z_j} U(z,t)\vert_{z=q(t)}\, \psi^{0,0}(x,t),\nonumber\\ 
&&
\end{eqnarray}
and
\begin{eqnarray}
&&\left( i\partial_{t}  + \frac{1}{2}\Delta -\frac{1}{\var}V_\var(x,t)-\frac{1}{\var} U(q(t)+\sqrt{\var}x,t) +\frac{1}{\var}U(q(t),t)+\frac{1}{\sqrt\var} \nabla U(q(t),t)\cdot x \right) \psi^{0,e_j}(x,t) = \nonumber\\
&&=\psi^{e_j,0}(x,t)\label{eq272}  
\end{eqnarray}
By virtue of the Duhamel formula we get:
\begin{eqnarray}
\label{eqDUH10}
 \psi^{e_j,0}(x,t) &=& P(t,0) \psi^{e_j,0}(x,0) +\int_0^t d\tau\, P(t,\tau)\, \left[\frac{1}{\var}\, \partial_{x_j} V_\var(x,\tau) \psi^{0,0}(x,\tau)\right]+\nonumber\\
&+ &\int_0^t d\tau\, P(t,\tau)\, \left[\frac{1}{\var}\, \partial_{x_j} U(q(\tau)+\sqrt{\var}x,\tau) \psi^{0,0}(x,\tau)-\frac{1}{\sqrt\var}\,  \pa_{z_j} U(q(\tau),\tau)\, \psi^{0,0}(x,\tau)\right] \nonumber  \\ 
&&
\end{eqnarray}
and
\begin{eqnarray}
\label{eqDUH01}
&& \psi^{0,e_j}(x,t) = P(t,0) \psi^{0,e_j}(x,0) +\int_0^t d\tau\, P(t,\tau)\, \left[\psi^{e_j,0}(x,\tau)\right].
\end{eqnarray}
Then, by recalling that $P(t,\tau)$ is  $L^2$-norm preserving, we find
\begin{eqnarray}
\label{eqDUH10L2}
 \left\|\psi^{e_j,0}(t)\right\|_{L^2} &\leq& \left\| \psi^{e_j,0}(0)\right\|_{L^2} +\int_0^t d\tau\, \left\|\frac{1}{\var}\, \partial_{x_j} V_\var(x,\tau) \psi^{0,0}(\tau)\right\|_{L^2}+\nonumber\\
&+& \int_0^t d\tau\, \left\|\left(\frac{1}{\var}\, \partial_{x_j} U(q(\tau)+\sqrt{\var}x,\tau) -\frac{1}{\sqrt\var}\,  \pa_{z_j} U(q(\tau),\tau) \right)\, \psi^{0,0}(\tau)\right\|_{L^2} \nonumber  \\ 
&&
\end{eqnarray}
and
\begin{eqnarray}
\label{eqDUH01L2}
&& \left\|\psi^{0,e_j}(t)\right\|_{L^2} \leq \left\| \psi^{0,e_j}(0)\right\|_{L^2} +\int_0^t d\tau\, \left\|\psi^{e_j,0}(\tau)\right\|_{L^2}.
\end{eqnarray}

\vskip 0.1 cm
Now, taking into account the terms involving $U$ in (\ref{eqDUH10L2}), we get
\begin{eqnarray}\label{unifBOUNDu}
&&\frac{1}{\var}\, \partial_{x_j} U(q(\tau)+\sqrt{\var}x,\tau)-\frac{1}{\sqrt\var}\,  \pa_{z_j} U(q(\tau),\tau)=\nonumber\\
&& = \frac{1}{\sqrt\var}\pa_{z_j} U(z,\tau)\vert_{z=q(\tau)+\sqrt{\var}x}-\frac{1}{\sqrt\var}\,  \pa_{z_j} U(z,\tau)\vert_{z=q(\tau)}=\nonumber\\
&&\nonumber\\
&&=\left[ \pa^2_{z_j} U(z,\tau)\vert_{z=q(t)+\sqrt{\delta}\, x}\right] x_j,\ \ \ \text{for some}\ \delta\in(0,\var),
\end{eqnarray}
therefore
\begin{eqnarray}\label{unifBOUNDu1}
&&\left\|\left(\frac{1}{\var}\, \partial_{x_j} U(q(\tau)+\sqrt{\var}x,\tau)-\frac{1}{\sqrt\var}\,  \pa_{z_j} U(q(\tau),\tau) \right)\psi^{0,0}(\tau)\right\|_{L^2}=\nonumber\\
&&\nonumber\\
&&= \left\|\left[ \pa^2_{z_j} U(z,\tau)\vert_{z=q(\tau)+\sqrt{\delta}\, x}\right] x_j\, \psi^{0,0}(\tau)\right\|_{L^2}=\left\|\left[ \pa^2_{z_j} U(z,\tau)\vert_{z=q(\tau)+\sqrt{\delta}\, x}\right]\psi^{0,e_j}(\tau)\right\|_{L^2}\leq\nonumber\\
&&\nonumber\\
&&\leq \sup_{\tau\in[0,t]}\left\|\pa^2 U(\cdot, \tau)\right\|_{L^\infty} \left\|\psi^{0,e_j}(\tau)\right\|_{L^2}.
\end{eqnarray}

On the other side, with regard to the term involving $V_\var$ in (\ref{eqDUH10L2}), we have
\begin{eqnarray}\label{unifBOUNDphi}
\left|\frac{1}{\var}\, \partial_{x_j} V_\var(x,\tau)\right|&=&
\left|\int d\eta\, \partial_{x_j} \frac{1}{\varepsilon} \phi( \sqrt{\var} |x-\eta|) \vert \psi^{0,0}(\eta, \tau)\vert^2\right| \leq\nonumber\\
&\leq &\int d\eta\,  \left|{ \frac{ \phi'(\sqrt{\var} |x-\eta|) }{\sqrt{\var}} }\right|\vert \psi^{0,0}(\eta, \tau)\vert^2 \leq L \int d\eta\,  |x-\eta|\vert \psi^{0,0}(\eta, \tau)\vert^2 ,
\end{eqnarray}
where $L$ is the global Lipschitz constant of $\phi'$ (i.e., the $L^\infty$-norm of $\phi''$) that is known to be finite since $\phi\in C_b^2(\R^d)$.
Then, by (\ref{unifBOUNDphi}) we get that
\begin{eqnarray}
\label{unifBOUNDphi2}
 \left\| \frac{1}{\var}\partial_{x_j}V_\var(x,\tau) \psi^{0,0}(\tau) \right\|_{L^2}^2&\leq& L^2 \int{ dx\int{d\eta|x-\eta| |\psi^{0,0}(\eta,\tau)|^2}\int{ d\eta'|x-\eta'| |\psi^{0,0}(\eta',\tau)|^2}|\psi^{0,0}(x,\tau)|^2 } \nonumber\\
%
%
&\leq& L^2 \int{|x|^2|\psi^{0,0}(x,\tau)|^2dx} + 3L^2 \left({ \int{|\eta| |\psi^{0,0}(\eta,\tau)|^2 d\eta} }\right)^2 \leqslant \nonumber\\
&\leq& L^2 ||\,\,|x|\psi^{0,0}(\tau)||^2_{L^2}+ 3L^2||\,\,|x|\psi^{0,0}(\tau)||^2_{L^2},
\end{eqnarray}
where we made use of the fact that $||\psi^{0,0}(\tau)||_{L^2}=||\psi^{0,0}(0)||_{L^2}=||a_0||_{L^2}=1$, for any time $\tau$.\\
At this point we observe that $||\,\,|x|\psi^{0,0}(\tau)||^2_{L^2} =||\,\,|x|a(\tau)||^2_{L^2}= \sum\limits_{j}||\psi^{0,e_j}(\tau)||^2_{L^2}$. So that, we have just proven that there exists a constant $C>0$ depending only on the  $L^\infty$-norm of the second derivative of $\phi$, such that
\begin{equation}
\label{eqzzoo}
 \left\| \frac{1}{\var}\partial_{x_j}V_\var(x,\tau) \psi^{0,0}(\tau) \right\|_{L^2} \leqslant C \sqrt{ \sum\limits_{j}||\psi^{0,e_j}(\tau)||_{L^2}^2}=||\psi^{0,1}(\tau)||_{L^2}.
\end{equation}

\vskip 0.1cm

Now, by using (\ref{unifBOUNDu1}) and (\ref{eqzzoo}) in (\ref{eqDUH10L2}), we obtain that
\begin{equation}
\label{eqzzoo007}
\begin{array}{c}
\left\|\psi^{e_j,0}(t)\right\|_{L^2} \leq \left\|\psi^{e_j,0}(0)\right\|_{L^2}+C\int\limits_{0}^{t} d\tau||\psi^{0,1}(\tau)||_{L^2} 
+ C\int\limits_{0}^{t} d\tau\left\|\psi^{0,e_j}(\tau)\right\|_{L^2},
\end{array}
\end{equation}
where $C$ is not the same constant of formula (\ref{eqzzoo}) - we denoted it by the same symbol just for the sake of simplicity - since here it is depending on $\phi$, as previously, but even on $U$ (through the $L^\infty$-norm of its second derivative, according to (\ref{unifBOUNDu1})).\\
Now after \eqref{unifBOUNDu1}, summing over $j=1,\dots, d$ in equations (\ref{eqzzoo007}) and (\ref{eqDUH01L2}) and then adding them, we get
\begin{equation}
\label{eqzzoo2007}
\begin{array}{c}
 ||\overrightarrow{\Psi}(t)||_{X_1}\leq  ||\overrightarrow{\Psi}(0)||_{X_1}+ C\int\limits_{0}^t d\tau\,  ||\overrightarrow{\Psi}(\tau)||_{X_1}.  \end{array}
\end{equation}

The conclusion follows by applying the Gronwall lemma, i.e.
\begin{equation}
\label{eqzzoo2008}
\begin{array}{c}
 ||\overrightarrow{\Psi}(t)||_{X_1}\leq  ||\overrightarrow{\Psi}(0)||_{X_1}e^{Ct}\leq M_1 e^{Ct},
 \end{array}
\end{equation}
where $M_1$ has been defined in (\ref{hypLEMMAa}).

\vskip 0.2cm

For $m\geq 2$, the previous inductive step from $m=1$ applies almost verbatim: first, by virtue of the Duhamel formula, we write the solution of equation (\ref{eqpab}) by using the propagator $P(t,\tau)$ associated with the time evolution on the left-hand side. Then, by using the $L^2$-control on $P(t,\tau)$, it only remains to show that the ``source terms'' appearing on the right-hand side  of (\ref{eqpab}) are suitably uniformly bounded in terms of $||\psi^{A,B}||_{L^2}$ or $||\overrightarrow{\Psi}||_{X_m}$. The way to do that is by using $||\psi^{A,B}||_{L^2}, |A|+|B|<m$ as constants now.


For example, let us look at the term involving the potential $V_\var$ on the right-hand side  of (\ref{eqpab}), i.e.
\begin{eqnarray}\label{new}
\frac{1}{\var}\sum\limits_{0 \leqslant l < A} { \prod\limits_{k=1}^d \binom{A_k}{l_k} \partial_x^{A-l}V_\var(x,t)  \psi^{l,B}(x,t)} 
&=&\frac{1}{\var}\sum\limits_{\substack{0 \leqslant l < A\\ |A-l|=1}} { \prod\limits_{k=1}^d \binom{A_k}{l_k} \partial_x^{A-l}V_\var(x,t)  \psi^{l,B}(x,t)} +\nonumber\\
&&+\frac{1}{\var}\sum\limits_{\substack{0 \leqslant l < A\\ |A-l|>1}} { \prod\limits_{k=1}^d \binom{A_k}{l_k} \partial_x^{A-l}V_\var(x,t)  \psi^{l,B}(x,t)}. 
\end{eqnarray}
The estimation for any of the terms in the last sum 
reads as:
\begin{equation}
\label{eqflerb}
\begin{array}{c}
|| \partial_x^{A-l}\int{\frac{1}{\var} \phi(\sqrt{\var}|x-\eta|)|\psi^{0,0}(\eta,t)|^2d\eta \,\, } \psi^{l,B}(x,t) ||_{L^2}^2 \leqslant \\ { } \\

\leqslant \left(\var^{\frac{|A-l|-2}{2}} ||\phi^{(A-l)}(x)||_{L^\infty}\right)^2
||  \int{ |x -\eta| |\psi^{0,0}(\eta,t)|^2d\eta }\,\, \psi^{l,B}(x,t)  ||_{L^2}^2 \leqslant \\ { } \\

\leqslant 2D ||\,\,|\eta| \psi^{0,0}(t)||_{L^2} ||\psi^{l,B}(t)||_{L^2} ||\,\,|x|\psi^{l,B}(t)||_{L^2}+ D||\,\,|\eta| \psi^{0,0}(t)||_{L^2}||\psi^{l,B}(t)||_{L^2}^2
+  D||\,\,|x|\psi^{l,B}(t)||_{L^2}^2=\\ { } \\

= 2 D||\psi^{0,1}(t)||_{L^2} ||\psi^{l,B}(t)||_{L^2} ||\psi^{l,B+1}(t)||_{L^2}+D||\psi^{0,1}(t)||_{L^2}^2||\psi^{l,B}(t)||_{L^2}^2+D||\psi^{l,B+1}(t)||_{L^2}^2,
\end{array}
\end{equation}
where 
$D$ is a constant only depending on $||\phi^{(A-l)}(x)||_{L^\infty}$ (that is finite under our assumptions since $ A - l \leq m$). Furthermore, it is clear that, by construction, we are guaranteed that the exponent $\frac{|A-l|-2}{2}$ for $\var$ is non negative.\\
On the other side, the estimate for any of the terms in the first sum on the right-hand side of (\ref{new}) is given by
\begin{equation}
\label{eqflerbL=1}
\begin{array}{c}
|| \partial_x^{A-l}\int{\frac{1}{\var} \phi(\sqrt{\var}|x-\eta|)|\psi^{0,0}(\eta,t)|^2d\eta \,\, } \psi^{l,B}(x,t) ||_{L^2}^2 
\leq \\ { } \\

\leqslant L
||  \int{ |x -\eta| |\psi^{0,0}(\eta,t)|^2d\eta }\,\, \psi^{l,B}(x,t)  ||_{L^2}^2 \leqslant \\ { } \\

\leqslant 2 L ||\,\,|\eta| \psi^{0,0}(t)||_{L^2} ||\psi^{l,B}(t)||_{L^2} ||\,\,|x|\psi^{l,B}(t)||_{L^2}+ L||\,\,|\eta| \psi^{0,0}(t)||_{L^2}||\psi^{l,B}(t)||_{L^2}^2
+  L||\,\,|x|\psi^{l,B}(t)||_{L^2}^2=\\ { } \\

= 2 L||\psi^{0,1}(t)||_{L^2} ||\psi^{l,B}(t)||_{L^2} ||\psi^{l,B+1}(t)||_{L^2}+L||\psi^{0,1}(t)||_{L^2}^2||\psi^{l,B}(t)||_{L^2}^2+L||\psi^{l,B+1}(t)||_{L^2}^2,\\ { } \\
\end{array}
\end{equation}
where $L$ is the global Lipshitz constant of $\phi'$ (see (\ref{unifBOUNDphi})), which is guaranteed to be finite since $\phi\in C_b^m(\R^d)$, with $m\geq 2$.

Now, by virtue of the estimate we proved for $m=1$ (see (\ref{eqzzoo2007})), from (\ref{eqflerb}) and (\ref{eqflerbL=1}) we find that
\begin{equation}
\label{eqflerbBIS}
\begin{array}{c}
|| \partial_x^{A-l}\int{\frac{1}{\var} \phi(\sqrt{\var}|x-\eta|)|\psi^{0,0}(\eta,t)|^2d\eta \,\, } \psi^{l,B}(x,t) ||_{L^2}^2 \leqslant \\ { } \\

\leq K M_1 e^{Ct} ||\psi^{l,B}(t)||_{L^2} ||\psi^{l,B+1}(t)||_{L^2}+ K M_1 e^{Ct}||\psi^{l,B}(t)||_{L^2}^2+K ||\psi^{l,B+1}(t)||_{L^2}^2,
\end{array}
\end{equation}
where $K=max\{D,L\}$ and
and we recall that $\vert l\vert+\vert B\vert\leq  \vert A \vert-1  + \vert B\vert\leq m-1$ and $\vert l\vert+\vert B\vert+1\leq \vert A \vert -1 + \vert B\vert+1\leq m $. Thus:
\begin{equation}
\label{eqflerbTRIS}
\begin{array}{c}
|| \partial_x^{A-l}\int{\frac{1}{\var} \phi(\sqrt{\var}|x-\eta|)|\psi^{0,0}(\eta,t)|^2d\eta \,\, } \psi^{l,B}(x,t) ||_{L^2}^2 \leq K(M_1 e^{Ct}+1) ||\overrightarrow{\Psi}(t)||_{X_m}^2.
\end{array}
\end{equation}

Concerning the terms involving the potential $U$ on the right-hand side  of (\ref{eqpab}), the idea is quite similar. In fact, we observe that
\begin{equation}
\label{eqPERu}
\begin{array}{c}
\frac{1}{\var}\sum\limits_{0 \leqslant l < A} { \prod\limits_{k=1}^d \binom{A_k}{l_k} \partial_x^{A-l}U(q(t)+\sqrt{\var}x,t)  \psi^{l,B}(x,t)}-\\
-\frac{1}{\sqrt{\var}}\sum\limits_{\substack{0 < l \leqslant A,\\\vert l\vert= 1}}
{ \prod\limits_{k=1}^d \binom{A_k}{l_k}\partial_x^{l}\left( \nabla U(q(t),t)\cdot x\right)\psi^{A-l,B}(x,t)}=  \\ { } \\

= \sum\limits_{\substack{0 < l \leqslant A,\\\vert l\vert = 1}} C_{A,l,B} { \left(\frac{1}{\var}{ \partial_x U(q(t)+\sqrt{\var}x,t) \psi^{A-l,B}(x,t)-\frac{1}{\sqrt{\var}}\partial_x\left( \nabla U(q(t),t)\cdot x\right)\psi^{A-l,B}(x,t)}\right) }+  \\ { } \\

+\frac{1}{\var}\sum\limits_{\substack{0 \leqslant l < A,\\ \vert A-l\vert >1}} { \prod\limits_{k=1}^d \binom{A_k}{l_k} \partial_x^{A-l}U(q(t)+\sqrt{\var}x,t)  \psi^{l,B}(x,t)},
\end{array}
\end{equation}
where we made a discrete change of variable $l\mapsto A-l$ in the first term of the left-hand side.

Now, with regard to first term of the right-hand side, the estimation that has to be used is exactly the one we did in (\ref{unifBOUNDu1}), thus one finds, $\forall \, l:\, |l|=1$
\begin{eqnarray}\label{unifBOUNDuFIN}
&&\left\|\frac{1}{\var}\partial_x U(q(t)+\sqrt{\var}x,t) \psi^{A-l,B}(t)-\frac{1}{\sqrt{\var}}\partial_x\left( \nabla U(q(t),t)\cdot x\right)\psi^{A-l,B}(t)\right\|_{L^2}\leq\nonumber\\
&&\leq \sup_{t}\left\|\pa^2 U(\cdot, t)\right\|_{L^\infty} \left\|\psi^{A,B}(t)\right\|_{L^2}\leq \sup_{t}\left\|\pa^2 U(\cdot, t)\right\|_{L^\infty}||\overrightarrow{\Psi}(t)||_{X_m}
\end{eqnarray}
(the adjustment for $l=0$ is obvious). \\
Now, for the last term in (\ref{eqPERu}) we have
\begin{eqnarray}\label{unifBOUNDuFIN1}
&&\left\|\frac{1}{\var}
\partial_x^{A-l}U(q(t)+\sqrt{\var}x,t)  \psi^{l,B}(x,t)\right\|_{L^2}\leq \var^{\frac{|A-l|-2}{2}} \sup_{t}||\pa^{(A-l)}U(\cdot, t)||_{L^\infty}\left\|\psi^{l,B}(t)\right\|_{L^2}\leq\nonumber\\
&&\leq \sup_{t}||\pa^{(m)}U(\cdot, t)||_{L^\infty}||\overrightarrow{\Psi}(t)||_{X_m},
\end{eqnarray}
where we used that $A-l\leq A\leq m$, $|A-l|-2\geq 0$ and $l+B<A+B\leq m$.\\

Similar (simpler, in fact) 
estimates can be shown for the other terms on the right-hand side  of (\ref{eqpab}).

\end{proof}

\vskip 0.3cm 
\begin{observation} \label{obsobsobsBETA} \emph{[Propagation of moments and derivatives  for $\beta_t(x)$]} $\beta_t(x)$ was defined in equations (\ref{machin}), (\ref{cxz}). Under the assumptions of Lemma \ref{lmmom}, 
regularity estimates for $\beta_t(x)$ analogous to Lemma \ref{lmmom} for $a(x,t)$ hold, i.e., for any $t>0$
$$
||x^B \pa_x^A \beta_t||_{L^2}\leqslant C_m(t) \sum\limits_{|A'|+|B'|\leqslant m}||x^{B'} \pa_x^{A'} a_0||_{L^2},\ \  \forall\  A,B\in \ \N^d:\  \vert A\vert + \vert B\vert\leq m.
$$
 \\\\\\\\
{\em {\bf Remarks:}
\begin{itemize}
\item The proof is in fact simpler with respect to the one of Lemma \ref{lmmom}: it can be checked easily that, due to the fact that we have to deal with harmonic potentials, the terms that arise from the differentiation of the potentials turn to be exactly of the form $x \beta_t(x)$ ($\sim||\overrightarrow{\Psi}_\beta(t)||_{X_1} $ if we denote by $\psi_\beta^{A,B}(x,t)$ the quantities $x^B\pa_x^A\beta_t(x)$ and we define $\overrightarrow{\Psi}_\beta(t)$ consistently), i.e., precisely the kind of objects we want to recover to apply the Gronwall Lemma (see the proof of Lemma \ref{lmmom}).\\
\item As a consequence of Observation \ref{obsobsobsBETA}, by Assumptions \ref{A1}, \ref{A2}, \ref{A3} we are guaranteed that, in particular, there exists a $\var$-independent constant $C>0$ depending on the $L^\infty$-norm of the second $x$-derivative of $U(x,t)$ and on $\vert \phi''(0)\vert$, such that
$$
\int dx\, \vert x\vert^{2} \vert \beta_t(x)\vert^2 <\left(\int dx\, \vert x\vert^{2} \vert a_0(x)\vert^2\right)\, e^{Ct}<\infty, \ \ \ \forall\ \ t.
$$
We remind that this is exactly what we need to make the proof of Theorem \ref{thrm1} work succesfully (see (\ref{MOMbetaPROD})).
%
\end{itemize}
}
\end{observation}


\vskip 0.3cm

\begin{observation} \label{obsobsobs}\emph{[Propagation of Moments and derivatives for $b_t(x)$]} Although apparently $b_t(x)$ solves a nonlinear equation,  it can be obtained as the solution of a linear Schr\"odinger equation with an harmonic potential whose coefficients are determined by the $L^2$-norm of the first moment of $\beta_t(x)$, by $\phi''(0)$ and $\nabla^2 U$ (see (\ref{Beqn2}) and Lemma \ref{14juillet}). \\ Therefore, as  a consequence of Observation \ref{obsobsobsBETA}, it follows that, as long as  $U \in C^1\left({ \R_t^+, C_b^{m}(\R_x^d)}\right)$ and $\phi''(0)$ is finite, we can get a result for $b_t(x)$ e.g. analogous to Lemma \ref{lmmom} for $a(x,t)$, i.e. 
$$
x^B \pa_x^A b_t \in L^2,\ \  \forall\  A,B\in \ \N^d:\  \vert A\vert + \vert B\vert\leq m\ \ \forall\ t,
$$
under the same assumption (\ref{hypLEMMAa}) on the (common) initial datum $a_0(x)$. 
\end{observation}

Note that, in particular, by Assumptions \ref{A1}, \ref{A2}, \ref{A3} we are guaranteed that  there exists a $\var$-independent constant $C_t>0$ depending on the $L^\infty$-norm of the second $x$-derivative of $U(x,t)$, on $\vert \phi''(0)\vert$ and on time $t$ (but finite for any $t$), such that
\begin{equation}\label{b3CONTROL}
\int dx\, \vert x\vert^{2m} \vert b_t(x)\vert^2 <C_t, \ \ \ \forall\ \ t,\ \ \ m\leq 3.
\end{equation}
We observe that (\ref{b3CONTROL}), for $m=2$, is exactly what we need to make the proof of Theorem \ref{thrm1} work succesfully (see (\ref{35}), (\ref{36}) and (\ref{RUcontrol})).\\

\vskip 0.3cm 

\section{Higher order approximations}


On the basis of the above results, it seems natural to ask whether it is possible to go beyond the 
$\sqrt{\var}$-approximation discussed previously (see \eqref{MAINclaim}) and to find higher order corrections 
$a^{(k)}_t(\mu)$ to the  amplitude  $a^{(0)}_t(\mu):=b_t(\mu)$ so that the right-hand side of \eqref{MAINclaim} gets of size of any
power of $\epsilon$, as this is the case for the linear Schr\"odinger equation \cite{tp1,tp2}.
Although we will not present all the (tedious) details of the construction, we claim that one can
determine a  semiclassical expansion 
\begin{equation}\label{eq: aEXP}
a^{\var}_t(\mu)=a^{(0)}_t(\mu)+\sqrt{\var} \, a^{(1)}_t(\mu)+\var\,  a^{(2)}_t(\mu)+\dots+\var^{k/2} a^{(k)}_t(\mu)+\dots,
\end{equation}
with
\begin{equation}\label{aEXPT0}
a^{(k)}_0(\mu)=\delta_{k,0}a_0(\mu),
\end{equation}
such that
\[
\Psi^\var(x,t)=e^{i\frac{\mathcal{L}(t)}{\var}+i\gamma(t)}\psi^{\beta_t}_{q(t)p(t)}
+O(\epsilon^\infty).
\]
In order to determine the equations governing the evolution for each coefficient $a^{(k)}_t(\mu)$ we need to look at the expansion for the potential terms appearing in (\ref{eq1sw}). With regard to the nonlinear part involving the pair interaction $\phi$, we get:
\begin{eqnarray}\label{phiEXP}
\frac{1}{\var}\left(\phi(\sqrt{\var}|\mu-\eta|)-\phi(0)\right)&=&\frac{ |\mu-\eta|}{\sqrt{\var}}\phi'(0)+\frac{ |\mu-\eta|^2}{2}\phi''(0)+ \frac{\sqrt{\var} |\mu-\eta|^3}{3!}\phi'''(0)+\nonumber\\
&+&\dots+ \frac{(\sqrt{\var})^k |\mu-\eta|^k}{k!}\phi^{(k)}(0)+\dots 
\end{eqnarray}
In Theorem \ref{thrm1} we were assuming $\phi\in C_b^3(\R)$. 
 Clearly, if we want to go to higher orders in the approximation we need more smoothness on $\phi$  and  on the external potential $U$.  Therefore, here and henceforth we assume:
\begin{equation}\label{hypPHICinfty}
\phi\in C_b^\infty(\R),\ \ \ \text{and}\ \  \ \phi \ \ \text{even}
\end{equation}
so that we have
\begin{eqnarray}\label{phiEXP2}
\frac{1}{\var}\left(\phi(\sqrt{\var}|\mu-\eta|)-\phi(0)\right)=\frac{ |\mu-\eta|^2}{2}\phi''(0)+\dots+ (\sqrt{\var})^{2n-1}\frac{ |\mu-\eta|^{2n}}{(2n)!}\phi^{(2n)}(0)+\dots\ \ n\geq 2
\end{eqnarray}
\begin{observation}\label{regPHI}
Assumption (\ref{hypPHICinfty}) is actually too strong if one wants to deal with an approximation up to a certain order $k$. 
\end{observation}
With regard to the linear terms in (\ref{eq1sw}) involving the external potential $U$, we get:
\begin{equation}\label{uEXP}
\begin{array}{c}
\frac{1}{\var}\left(U(q(t)+\sqrt\var\mu,t)-U(q(t),t)-\sqrt\var \nabla U(q(t),t)\cdot \mu\right) = \\ { } \\

= <\mu,\frac{\nabla^2 U(q(t),t)}2\mu> + 
\dots + (\sqrt{\var})^{n-1}\frac{\nabla^n U(q(t),t)}{n!}\cdot \mu^n+\dots\nonumber\\
\end{array}
\end{equation}
where of course $n\geqslant 3$. Here we are using the notation
\begin{equation}\label{not}
\nabla^n U\cdot \mu^n=\sum_{\substack{\alpha_1 \dots \alpha_d: \\ \sum \alpha_i=n }} \frac {\partial^n U}{\partial x_1^{\alpha_1} \dots \partial x_d^{\alpha_d}}\, \,   \mu^{\alpha_1} \dots \mu ^{\alpha_d}.
\end{equation}

Analogously to what we observed for the pair-interaction $\phi$, we need more smoothness for $U$, so that here and henceforth we require:
\begin{equation}\label{hypUCinfty}
U\in C^1(\R^+_t, C_b^\infty(\R_x^d)).
\end{equation}

Now, inserting (\ref{eq: aEXP}), (\ref{phiEXP2}) and (\ref{uEXP}) in (\ref{eq1sw}) 
we readily arrive to a sequence of problems for the coefficients $ a^{(k)}_t(\mu)$ of the expansion (\ref{eq: aEXP}). For $k=0$ we 
obviously find:
\begin{equation}\label{eqk=0}
\left\{
\begin{aligned}
&\left(i\pa_t+\frac{\Delta_\mu}{2}+ \frac{\phi''(0)}{2}\int d\eta\, \vert \mu-\eta\vert^2\vert a^{(0)}_t(\eta)\vert^2 + <\mu,\frac{ \nabla^2 U(q(t),t)}{2}\mu> \right)a^{(0)}_t(\mu)=0,\\
&a^{(0)}_0(\mu)=a_0^\var(\mu),
\end{aligned}
\right.
\end{equation}
namely, the initial value problem that we had for $b_t(\mu)$ in the previous sections (see (\ref{truc})). 
Then, for $k=1$, we find:
\begin{equation}\label{eqk=1}
\left\{
\begin{aligned}
&\left(i\pa_t+\frac{\Delta_\mu}{2}+ \frac{\phi''(0)}{2}\int d\eta\, \vert \mu-\eta\vert^2\vert a^{(0)}_t(\eta)\vert^2 + <\mu,\frac{ \nabla^2 U(q(t),t)}{2}\mu>\right)a^{(1)}_t(\mu)=\\
&=\frac{\phi''(0)}{2}\left(\int d\eta\, \vert \mu-\eta\vert^2 2\Re[\overline{a}^{(0)}_t(\eta)a^{(1)}_t(\eta)]\right) a^{(0)}_t(\mu)+\frac{\nabla^3 U(q(t),t)}{3!}\cdot \mu^3\,  a^{(0)}_t(\mu),\\
&a^{(1)}_0(\mu)=0.
\end{aligned}
\right.
\end{equation}
This is a linear initial value problem where the left-hand side is known to have a unique well-defined $L^2$-propagator $P^{(0)}(t)$ due to the existence and uniqueness in $L^2$ of the solution $a^{(0)}_t(\mu)$ of the zero-order initial value problem (\ref{eqk=0}) and to the $L^2$-control on its first moment (see Observation \ref{obsobsobs} and (\ref{b3CONTROL})). Then, it is easy to see that, writing the solution $a^{(1)}_t(\mu)$ through the Duhamel formula (with ``leading'' propagator $P^{(0)}(t)$), the well-posedeness in $L^2$ for (\ref{eqk=1}) is guaranteed by the $L^2$-control on the source term $\frac{\nabla^3 U(q(t),t)}{3!}\cdot \mu^3\,  a^{(0)}_t(\mu)$ (which is achieved thanks to the smoothness of $U$ and to the $L^2$-control on the third moment of $a^{(0)}_t(\mu)$ - see (\ref{b3CONTROL})) and by the following estimate:
\begin{eqnarray}\label{source1}
&&\left\|\frac{\phi''(0)}{2}\left(\int d\eta\, \vert \mu-\eta\vert^2 2\Re[\overline{a}^{(0)}_t(\eta)a^{(1)}_t(\eta)]\right) a^{(0)}_t(\mu)\right\|_{L^2}^2= \nonumber\\
&&=
\left(\frac{\phi''(0)}{2}\right)^2\int d\mu\, \int d\eta \vert \mu-\eta\vert^2\overline{a}^{(0)}_t(\eta)a^{(1)}_t(\eta)\int d\eta' \vert \mu-\eta'\vert^2\overline{a}^{(0)}_t(\eta')a^{(1)}_t(\eta')\, \vert a^{(0)}_t(\mu)\vert^2+\nonumber\\
&&+\left(\frac{\phi''(0)}{2}\right)^2\int d\mu\, \int d\eta \vert \mu-\eta\vert^2a^{(0)}_t(\eta)\overline{a}^{(1)}_t(\eta)\int d\eta' \vert \mu-\eta'\vert^2 a^{(0)}_t(\eta')\overline{a}^{(1)}_t(\eta')\, \vert a^{(0)}_t(\mu)\vert^2\leq\nonumber\\
&&\leq C_{a^{(0)}_t} \left\| a^{(1)}_t\right\|_{L^2}^2.
\end{eqnarray}
Here $C_{a^{(0)}_t}>0$ is a constant that only depends on the moments of $a^{(0)}_t(\mu)$ up to  order $3$,  depending on the potentials $U$ (trought the $L^\infty$-norm of its second derivative) and $\phi$ (through the quantity $\vert \phi''(0)\vert$) and on the initial derivatives and moments of $a^{(0)}_t(\mu)$ up to the order $3$ (see Observation \ref{obsobsobs}), that, as in the previous sections, we assume to be finite.
By virtue of (\ref{source1}) and the $L^2$-control on the term involving $U$ in (\ref{eqk=1}), the Duhamel formula and the Gronwall lemma allow to conclude that 
\begin{equation}\label{a1L2}
a^{(1)}_t \in L^2(\R^d),\ \ \forall\ t.
\end{equation}
Moreover, following the same lines of the proofs presented and discussed in the previous sections (see Lemma \ref{lmmom}, Observation \ref{obsobsobs} and subsequent remarks), it can be easily checked that by assuming enough regularity for the (zero-order) ``full'' initial datum $a_0^{(0)}(\mu)=a_0(\mu)$, in such a way that we control in $L^2$
a sufficiently high number of moments and derivatives,
we can control the derivatives and moments of $a^{(1)}_t(\mu)$ up to any fixed order $m$, i.e:
\begin{equation}\label{a1moments}
x^B\pa_x^A a^{(1)}_t \in L^2(\R^d),\ \ \forall\ t,\ \ \ \ \forall\ \ A,B\in \N^d: \vert A\vert+\vert B\vert\leq m.
\end{equation}
This will be crucial to go on with the higher orders dynamics because, for example, the equation for the second coefficient $a^{(2)}_t(\mu)$ is
\begin{equation}\label{eqk=2}
\left\{
\begin{aligned}
&\left(i\pa_t+\frac{\Delta_\mu}{2}+ \frac{\phi''(0)}{2}\int d\eta\, \vert \mu-\eta\vert^2\vert a^{(0)}_t(\eta)\vert^2 + <\mu,\frac{ \nabla^2 U(q(t),t)}{2}\mu>\right)a^{(2)}_t(\mu)=\\
&=\frac{\phi''(0)}{2}\left(\int d\eta\, \vert \mu-\eta\vert^2 2\Re[\overline{a}^{(0)}_t(\eta)a^{(2)}_t(\eta)]\right) a^{(0)}_t(\mu)
+\frac{\nabla^4 U(q(t),t)}{4!}\cdot \mu^4\,  a^{(0)}_t(\mu)+\\
&+\frac{\phi^{(4)}(0)}{4!}\left(\int d\eta\, \vert \mu-\eta\vert^4\vert a^{(0)}_t(\eta)\vert^2\right)a^{(0)}_t(\mu)
+\frac{\phi''(0)}{2}\left(\int d\eta\, \vert \mu-\eta\vert^2\vert a^{(1)}_t(\eta)\vert^2\right)a^{(0)}_t(\mu)+\\
&+\frac{\phi''(0)}{2}\left(\int d\eta\, \vert \mu-\eta\vert^2 2\Re[\overline{a}^{(0)}_t(\eta)a^{(1)}_t(\eta)]\right) a^{(1)}_t(\mu)+\frac{\nabla^3 U(q(t),t)}{3!}\cdot \mu^3\,  a^{(1)}_t(\mu).\\
&a^{(2)}_0(\mu)=0,
\end{aligned}
\right.
\end{equation}
So, again, as for the case $k=1$, we obtained a linear initial value problem where the propagator associated with the left-hand side is $P^{(0)}(t)$, that is known to be uniquely well-defined in $L^2$. Then, as before, the solution $a^{(2)}_t(\mu)$ can be written through the Duhamel formula, applying the propagator $P^{(0)}(t)$ to the term $\frac{\phi''(0)}{2}\left(\int d\eta\, \vert \mu-\eta\vert^2 2\Re[\overline{a}^{(0)}_t(\eta)a^{(2)}_t(\eta)]\right) a^{(0)}_t(\mu)$ and to the various source terms in (\ref{eqk=2}). The term which is linear in $a^{(2)}_t(\mu)$ is estimated as in (\ref{source1}) while the source terms are controlled in $L^2$ by virtue of the control on moments and derivatives of $a^{(0)}_t(\mu)$ and $a^{(1)}_t(\mu)$.
In the end, by using the Gronwall lemma, we get
\begin{equation}\label{a2L2}
a^{(2)}_t \in L^2(\R^d),\ \ \forall\ t.
\end{equation}
and, moreover, by assuming a sufficiently high number of moments and derivatives of the (zero-order) ``full'' initial datum $a_0^{(0)}(\mu)=a_0(\mu)$ to be controlled in $L^2$, we can control as well the derivatives and moments of $a^{(2)}_t(\mu)$ up to any fixed order $m$, i.e:
\begin{equation}\label{a2moments}
x^B\pa_x^A a^{(2)}_t \in L^2(\R^d),\ \ \forall\ t,\ \ \ \ \forall\ \ A,B\in \N^d: \vert A\vert+\vert B\vert\leq m.
\end{equation}

At this point it is clear how to proceed in general.  The equation for $a^{(k)}_t(\mu)$ is a linear Schr\"odinger equation with a source term involving the coefficients
$a^{(n)}_t(\mu)$ with $n<k$, which have been estimated by the previous steps. The $L^2$-control of $a^{(k)}_t(\mu)$ follows by the $L^2$-control on a sufficiently
high number of moments and derivatives of $a^{(n)}_t(\mu)$ with $n<k$.

\vskip 0.4cm
\noindent {\bf Acknowledgments.} We would like to thank R. Carles for pointing out that an extra hypothesis was needed, and a mistake in the remainder which appears in the main theorem.


\begin{thebibliography}{99}
\bibitem{appp1} A. Athanassoulis, T. Paul, F. Pezzotti  \& M. Pulvirenti, \emph{Strong semiclassical approximation of Wigner
functions for the Hartree dynamics}, 	arXiv:1009.0470v1 [math-ph].

\bibitem{Bel2}
V. V. Belov, M. F. Kondratieva \& E. I. Smirnova, \emph{Semiclassical Soliton-Type Solutions of the Hartree Equation}. Doklady math {\bf 76} No. 2 775-779 (2007).
\bibitem{carlito} R. Carles \& C. Fermanian-Kammerer, 
\emph{Nonlinear coherent states and Eherenfest time for Schr\"odinger equation}, arXiv:0912.1939v1.

\bibitem{CW}
T. Cazenave \& F. Weissler, \emph{The Cauchy problem for the nonlinear Schr\"odinger equation in $H^1$}. \ manuscripta mathematica {\bf 61} 477-494 (1988).

\bibitem{GV}
J. Ginibre \& G. Velo, \emph{On a class of non linear Schr\"odinger equations with non local interactions}. Mathematische Zeitschrift {\bf 170} No. 2 109-136 (1980).

\bibitem{HAG1} 
G.A. Hagedorn,\ \emph{Semiclassical quantum mechanics. I.  The $\hbar\to 0$ limit for coherent states}.\ 
Comm. Math. Phys. {\bf 71}, No. 1  (1980).

\bibitem{HAG} 
G.A. Hagedorn, \emph{ Raising and Lowering Operators for Semiclassical Wave Packets}. Ann. Phys. {\bf 269}, 77--104 (1998).

\bibitem{HEPP} 
K. Hepp,\ \emph{The classical limit for quantum mechanical correlation functions}.\ Commun. Math. Phys.  {\bf 35} (1974).


\bibitem{LP}
P.-L. Lions \& T. Paul, \emph{Sur les mesures de Wigner}. Rev. Mat. Iberoamericana {\bf 9} No. 3 553-618 (1993).


\bibitem{Bel1}
A. Lisok, A. Yu. Trifnov \& A. V. Shapovalov, \emph{The evolution operator of the Hartree-type equation with a quadratic potential}.  Journal of Physics A {37} 4535 (2004). 




\bibitem{tp1}T. Paul \emph{Semiclassical methods with an emphasis on coherent states}, Tutorial Lectures,
Proceedings of the conference "Quasiclassical methods", B. Simon and J. Rauch, eds., IMA Series, Springer Verlag (1997).

\bibitem{tp2}T. Paul \emph{\'Echelles de temps pour l'\'evolution quantique \`a petite constante de Planck}, S\'eminaire X-EDP 2007-2008, Publications
de l'\'Ecole Polytechnique (2008).

\bibitem{SR}
 M. Reed \& B. Simon \emph{Methods of modern mathematical physics}. Vol. II, Academic Press (1975).

\end{thebibliography}
\end{document}